# Integral Formulas for Electrically Charged Space Regions - Theory and Application

*Miron Jean Cristea*

*"Politehnica" University of Bucharest, Bd. Iuliu Maniu 1-3, Office B-111, Sector 6, Romania*

E-mail: *mcris@lydo.org*

**Abstract.** Based on Gauss's law for the electric field, new integral formulas are deduced. Although their applications are not limited within the physics realm, an application is also presented, for the sake of practicability, specifically in the area of semiconductor junctions.

## 1 Introduction

The Gauss's law connects the electric field with the space charge density [1]:

$$\frac{dE}{dx} = \frac{\rho(x)}{\varepsilon} \qquad (1)$$

where $E$ is the electric field, $\rho$ is the electric space charge density and $\varepsilon$ the electrical permittivity.

The electric field is also connected with the electric potential along the field lines:

$$E = -\frac{du}{dx} \qquad (2)$$

Here $u$ is the electric potential.

Usually, these equations are merged in what is called the Poisson equation:

$$\frac{d^2u}{dx^2} = -\frac{\rho(x)}{\varepsilon} \qquad (3)$$

This equation is widely used, although it is not always very useful, because it requires a double integration in order to compute the potential distribution. This is not always possible, due to the particular mathematical form of the space charge density *ρ(x)*.

Much more useful would prove an integral formula deduced from eqns. (1) and (2). Such integral formulas are presented in the following. It is expected -and proved- they allow the resolution of previously unsolved problems.

## 2 Theory

By writing (1) as

$$dE = \frac{\rho(x)}{\varepsilon} dx \qquad (4)$$

multiplying the equation by $x$, and taking into account that

$$d(xE) = xdE + Edx \qquad (5)$$

the next equation is obtained:

$$d(xE) - Edx = \frac{\rho(x)}{\varepsilon} xdx \qquad (6)$$

The integration of (6) over the space charge region (SCR) gives

$$\int_{SCR} \frac{x\rho(x)}{\varepsilon} dx = \int_{SCR} d(xE) - \int_{SCR} Edx \qquad (7)$$

and taking (2) into account, then

$$\int_{SCR} \frac{x\rho(x)}{\varepsilon} dx = \int_{SCR} d(xE) + \int_{SCR} du \qquad (8)$$

Both terms in the right hand of this equation are perfect integrals. There are two particular cases:

1. the electric field is zero at both ends of the SCR, or
2. the electric field is zero at only one end of the SCR and the origin of the coordinates ($x=0$) is chosen at the other end, having there a finite value for the electric field.

In these cases the first term in the right hand of (8) is equal with zero, and the next formula is obtained:

$$\int_{SCR} \frac{x\rho(x)}{\varepsilon} dx = V_{SCR} \qquad (9)$$

where $V_{SCR}$ is the total voltage drop across the space charge region. In the general case, the form of the formula is:

$$\int_{SCR} \frac{x\rho(x)}{\varepsilon} dx = V_{SCR} - x_1 E(x_1) + x_2 E(x_2) \qquad (10)$$

where $x_1$ and $x_2$ are the spatial limits of the SCR ( i.e. SCR$\in [x_1\ x_2]$ ).

The applications of this formula extend throughout the physics, particularly in the electromagnetic field theory. One of these applications is presented in the following.

## 3 Application to semiconductor junctions

Particularization to semiconductor junctions leads to

$$\int_{SCR} \frac{x\rho(x)}{\varepsilon} dx = V_{bi} - V_F \qquad (11)$$

where $V_{bi}$ is the built-in voltage of the junction and $V_F$ is the externally applied forward bias, to be replaced with - $V_R$ in the case of an external reverse bias [2]. Practical applications of this formula include calculation of the depletion region width and barrier capacitance of diffused semiconductor junctions having Gaussian doping profiles. A quadratic exponential form of the doping profile - e.g. $C_0 \exp(-x^2/L^2)$ - characterizes these junctions, hence the distribution of the space charge density function. Because of that, equations (1) and (3) cannot be analytically integrated. But as it is noticeable in formula (11), the space charge function is multiplied by $x$, which makes possible to integrate it.

## 4 Conclusion

In this work, based on Gauss's law for the electric field, new integral formulas were deduced and one of the possible applications, namely in the area of semiconductor junctions, was outlined.